\documentclass[twocolumn,prd,floatfix,amsmath,nofootinbib,amssymb,floatfix]{revtex4}
\usepackage{graphicx,color,dcolumn,booktabs,bm}
\usepackage{longtable,lscape}
\usepackage{txfonts}
\usepackage{overpic}
\usepackage{amssymb}
\usepackage{indentfirst}
\usepackage{feynmf}
\usepackage{slashed}
\usepackage{cases}
\usepackage{multirow}
\usepackage[colorlinks,
            citecolor=blue,
            anchorcolor=red,
            menucolor=red,
            linkcolor=red,
            filecolor=red,
            runcolor=red,
            urlcolor=blue,
            frenchlinks=red]{hyperref}
\usepackage{epstopdf}
\usepackage{bm}
\usepackage{float}
\begin{document}
%******************************************************************
\title{$J/\psi$ photoproduction near threshold and signals for the hidden charm pentaquarks }
\author{Ming-Xiao Duan$^{1}$~\footnote{E-mail: duanmx@ihep.ac.cn}, Chang Gong$^{2}$~\footnote{E-mail: gongchang@ccnu.edu.cn}, Lin Qiu$^{1,3}$~\footnote{E-mail: qiulin@ihep.ac.cn}, and
Qiang Zhao$^{1,3,4}$~\footnote{E-mail: zhaoq@ihep.ac.cn}}

\affiliation{
$^1$ Institute of High Energy Physics, Chinese Academy of Sciences, Beijing 100049, China \\
$^2$ Key Laboratory of Quark and Lepton Physics (MOE) and Institute of Particle Physics, Central China Normal University, Wuhan 430079, China\\
$^3$ University of Chinese Academy of Sciences, Beijing 100049, China \\
$^4$ Center for High Energy Physics, Henan Academy of Sciences, Zhengzhou 450046, China}

\begin{abstract}
We study the $J/\psi$ photoproduction $\gamma p\to J/\psi p$ near threshold to investigate the role played by the open charm channels and possible signals from the hidden charm $P_c$ pentaquark states. With the diffractive mechanism described by a Pomeron exchange model extrapolated from high energies to the lower energy region, it shows that the differential cross sections compared with the recent results from the GlueX Collaboration have apparent deficits in the large scattering angles. The inclusion of the open charm channels and intermediate $P_c$ states can significantly improve the descriptions of the differential cross section data, in particular, at the energy regions of the $\Lambda_c\bar{D}^{(*)}$ thresholds. This can explain the structures observed by the GlueX Collaboration at the $\Lambda_c\bar{D}^{(*)}$ thresholds as the open charm CUSP effects. Given that these $P_c$ pentaquark states are hadronic molecules dynamically generated by the $\Sigma_c\bar{D}^{(*)}$, we find that the production of these $P_c$ states should be suppressed at leading order since their couplings to both $\gamma p$ and $J/\psi p$ are through loop transitions. This can explain why no obvious signals for the $P_c$ states are observed by the present datasets. In order to further disentangling the role played by $s$-channel mechanisms, we also investigate the polarized beam asymmetry which shows sensitivities to the open charm threshold effects and interferences from the $P_c$ productions. Experimental measurement of this observable at the Jefferson Laboratory is strongly encouraged.
\end{abstract}

\pacs{11.55.Fv, 12.40.Yx ,14.40.Gx}
\maketitle

%******************************************************************
\section{introduction}\label{sec1}
%******************************************************************
The discovery of the hidden charm $P_c$ pentaquark states by the LHCb Collaboration~\cite{LHCb:2015yax,LHCb:2019kea} has initiated extensive interests and studies in both experiment and theory. While the observed masses are located near the $\Sigma_c^{(*)}\bar{D}^{(*)}$ thresholds, these observed states seem to be consistent with the predicted $\Sigma_c^{(*)}\bar{D}^{(*)}$ molecules based on the charmed baryon and $\bar{D}^{(*)}$ interactions~\cite{Wu:2010vk, Yang:2011wz, Wang:2011rga, Wu:2012md}. Such a molecular picture has been broadly explored~\cite{Chen:2019bip, Chen:2019asm, Liu:2019tjn, He:2019ify, Guo:2019kdc, Xiao:2019mvs, Xiao:2019aya, Meng:2019ilv, Yamaguchi:2019seo, Du:2019pij, Wang:2019ato}, and various possible explanations concerning the structures and production mechanisms have been proposed in the literature (see recent reviews by Refs.~\cite{Zhao:2016akg, Guo:2017jvc, Chen:2016qju,Olsen:2017bmm,Esposito:2016noz,Lebed:2016hpi,Brambilla:2019esw,Liu:2019zoy,Ali:2017jda,Karliner:2017qhf,Chen:2022asf}).

Although the LHCb observations have provided strong evidences for the existence of the hidden charm pentaquark states as hadronic molecules, further confirmations of their signals are still needed. In Refs.~\cite{Wang:2015jsa} it is proposed that the $J/\psi$ photoproduction may provide a unique probe for the study of the $P_c$ states. Apart from the dominance of the diffractive transitions in $\gamma p\to J/\psi p$ at small scattering angles, the production of $P_c$ states will be through the $s$-channel transition. Thus, they will enhance the differential cross sections at middle and backward angles. This feature is extensive explored in Refs.~\cite{Gryniuk:2016mpk,Meziani:2016lhg,Paryev:2018fyv,Wang:2019krd,Wu:2019adv}.

In Ref.~\cite{GlueX:2019mkq} the GlueX Collaboration published their first search for the $P_c$ states in $\gamma p\to J/\psi p$ near threshold. Although the total cross section data did not show clear structures near threshold, the large uncertainties would not eliminate the possible contributions from the $P_c$ productions in the $s$ channel. In 2023, the GlueX Collaboration reported the updated total cross section measurement in association with the differential cross sections~\cite{GlueX:2023pev}. With the photon beam energy covers a range from 8.2 GeV to 11.2 GeV, the thresholds for both the $\Lambda_c\bar{D}^{(*)}$ and $\Sigma_c\bar{D}^{(*)}$ thresholds and the $P_c$ states can be accessed. Interestingly, in the total cross section two structures can be recognized at the $\Lambda_c\bar{D}$ and $\Lambda_c\bar{D}^{(*)}$ thresholds, but no signs for the $P_c$ states can be identified in the spectrum based on the present statistics. Regarding the differential cross sections, one sees the dominance of the forward-angle diffractive contributions. Meanwhile, flatten behaviors appear at large scattering angles which seem to be deviated from the diffractive mechanism.

In Ref.~\cite{Duan:2023dky} a ``triangle singularity (TS)" mechanism was identified in the $P_c$ decays into $J/\psi p\pi$, where the invariant mass spectrum of $J/\psi p$ is predicted to have narrow peaks to appear at the mass thresholds of $\Lambda_c\bar{D}^{(*)}$. This unique feature can be regarded as a special phenomenon arising from the molecular picture for the $P_c$ states of which the decays will go through the rescattering of the constituent hadrons and produce signature effects. While it is pointed out in Ref.~\cite{Duan:2023dky} that the $\Lambda_c\bar{D}^{(*)}$ thresholds may manifest themselves in association with the $P_c$ production in the near-threshold scatterings of $\Sigma_c^{(*)}$ and $\bar{D}^{(*)}$, they may also produce peculiar effects in $J/\psi$ photoproduction and guide our understanding of the nature of these $P_c$ states. This motivates us to revisit the $J/\psi$ photoproduction near threshold by taking into account the experimental constraints from the GlueX measurement~\cite{GlueX:2019mkq,Duan:2023dky}.

The $J/\psi$ photoproduction process is a long-lived topic in the hadron physics~\cite{Donnachie:1984xq, Donnachie:1987pu, Camerini:1975cy, Gittelman:1975ix, Shambroom:1982qj, ZEUS:1995kab, ZEUS:2002wfj}. The Pomeron-exchange scheme was successfully applied in the high-energy diffractive processes~\cite{Pichowsky:1996jx, Pichowsky:1996tn, Laget:1994ba, Zhao:1999af, Wu:2019adv, Lee:2022ymp, Tang:2024pky}. The Pomeron behaves like a $C=+1$ isoscalar photon, and different from the $t$-channel meson exchanges. The explicit form and parameters of the Pomeron exchange are determined from the study at the high energy and low momentum transfer region, where the diffractive scattering governs the reaction mechanism in the photoproduction. For the photoproduction of $J/\psi$ the Pomeron condition, $s>> |t|$ can still be fulfilled near threshold. Therefore, the diffractive contributions can still be described by the Pomeron exchange with an extrapolation from high energies. Our attention is paid to the energy region covered by the GlueX experiment, i.e. 8.2 GeV$<E_\gamma<$11.2 GeV. In particular, we will examine the role played by the open channel thresholds of $\Lambda_c\bar{D}^{(*)}$ and the $P_c$ states at large scattering angles in the differential cross section. Note that due to the dominance of the diffractive contributions the total cross section would not be a sensitive observable for the $s$-channel transitions.

It should be addressed that the $J/\psi$ photoproduction at JLab has a special advantage for the measurement of the linearly polarized photon beam asymmetry~\cite{GlueX:2020idb}. Investigation of the polarization observables for vector meson photoproductions can be found in the literature~\cite{Conzett:1994rg,Pichowsky:1994gh,Zhao:1998fn,Zhao:1998rt, Zhao:2005vh}. At large scattering angles the interferences between the dropping diffractive contribution and $s$-channel amplitude can be amplified and may shed light on the transition mechanisms.

Our paper is organized as follows. Following the introduction, the formalism for the $J/\psi$ photoproduction mechanism is presented in Sec.~\ref{sec2}. The numerical results for the total cross sections, differential cross sections, and the polarized beam asymmetry are given in Sec.~\ref{sec3}. A summary is presented in Sec.~\ref{sec4}.

%***********************************************************************
\section{Formalism}\label{sec2}
\begin{center}
\begin{figure}[htbp]
\includegraphics[width=8.4cm,keepaspectratio]{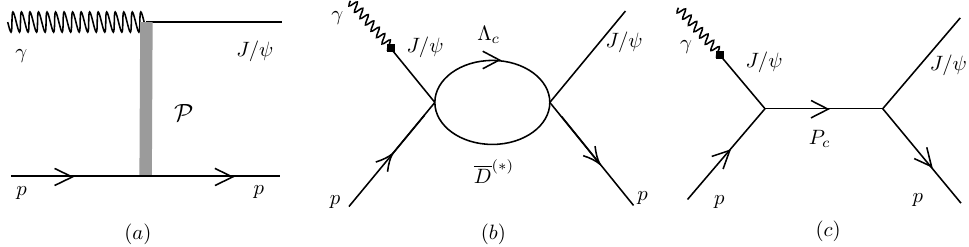}
\caption{Feynman diagrams of (a) the Pomeron exchange, (b) intermediate $\Lambda_c\bar{D}^{(*)}$ loops, and (c) the $s$-channel $P_c$ production.}
\label{f1}
\end{figure}
\end{center}

%***********************************************************************
In Fig.~\ref{f1} the main reaction mechanism for $\gamma p\to J/\psi p$ in the energy region near threshold is illustrated. Namely, the diffractive contribution is described by the Pomeron exchange mechanism (Fig.~\ref{f1}(a)) and the $s$-channel open-charm contributions and $P_c$ productions are described by the $\Lambda_c\bar{D}^{(*)}$ loops and $s$-channel pole structures (Fig.~\ref{f1}(b, c)), respectively. As follows, we will give details for extracting the transition amplitudes for each process.

\subsection{Pomeron exchange}
For the exclusive reaction $\gamma p\to J/\psi p$ the diffractive mechanism needs a proper prescription. As studied in the literature~\cite{Pichowsky:1996jx, Pichowsky:1996tn, Laget:1994ba, Zhao:1999af, Wu:2019adv, Lee:2022ymp, Tang:2024pky} the Pomeron exchange mechanism, which generally applies to the high energy processes, should still play a role in the kinematic region near threshold.

\begin{center}
\begin{figure}[htbp]
\includegraphics[width=6.4cm,height=3.5cm]{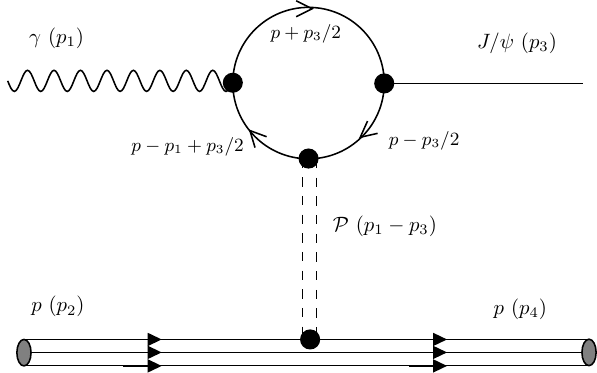}
\caption{The Pomeron exchange diagram at the quark level.}
\label{f2}
\end{figure}
\end{center}

In the Fig.\ref{f2}, the Feynman diagram of the Pomeron exchange is shown. As an effective phenomenological mechanism for the $t$-channel multi-soft gluon exchanges, the Pomeron behaves like a $C=$+1 isoscalar photon. The interaction between the Pomeron and a quark or anti-quark contributes a Lorentz index with the Dirac matrix $\gamma_\alpha$, and the Pomeron-nucleon coupling vertex can be expressed as:
\begin{equation}\label{eq2}
\begin{split}
F_\alpha(t)=3\beta_0\gamma_\alpha f(t),\\
\end{split}
\end{equation}
where $\beta_0=2.07~{\rm GeV}^{-1}$ represents the coupling strength between the Pomeron and a light constitute quark in the proton, and $t$ is the squared momentum of the Pomeron. Function $f(t)$ is a form factor taken from the isoscalar nucleon electromagnetic form factor, which has the following expression:
\begin{equation}\label{eq3}
\begin{split}
f(t)=\frac{4M_{\rm p}^2-2.8t}{(4M_{\rm p}^2-t)(1-t/0.7)^2}.\\
\end{split}
\end{equation}
Here, the $M_{\rm p}$ denotes the proton mass.

For the $\gamma\mathcal{P}J/\psi$ vertex, the initial photon, the exchanged Pomeron and the final $J/\psi$ states are linked by a charm quark loop. At the vertex of $J/\psi$ and charm quark in the loop, the on-shell approximation, which originates from the eikonal approximation for the exchanged soft gluons, is employed.  The Lorentz structure of the $\gamma\mathcal{P}J/\psi$ vertex is finally determined as:
\begin{equation}\label{eq4}
T^{\alpha,\mu\nu}=(p_1+p_3)^\alpha g^{\mu\nu}-2p_1^\nu g^{\alpha\mu},\\
\end{equation}
where $p_1$ and $p_3$ represent the momentum of the photon and $J/\psi$ in the $\gamma\mathcal{P}J/\psi$ vertex. With the Regge trajectory of the Pomeron, the exchanged Pomeron in the amplitude can be written as
\begin{equation}\label{eq5}
\mathcal{G}_\mathcal{P}(s,t)=-i(\alpha^\prime s)^{\alpha(t)-1},\\
\end{equation}
where $\alpha(t)=1+\epsilon+\alpha^\prime t$ is the Regge trajectory of the Pomeron.

With the above consideration, the amplitude of the Pomeron interaction has the following form:
\begin{equation}\label{eq6}
\begin{split}
\mathcal{M}^{\mathcal{P}}=&\frac{eM_{J/\psi}^2}{f_{J/\psi}}\varepsilon_{\gamma\nu}\left[2\beta_cT^{\alpha,\mu\nu}\frac{4\mu_0}{(M_{J/\psi}^2-t)(2\mu_0^2+M_{J/\psi}^2-t)}\right]\varepsilon_{J/\psi\mu}^*\\
&\times\bar{u}(p_4)F_\alpha(t)u(p_2)\left[-i(\alpha^\prime s)^{\alpha-1}\right].\\
\end{split}
\end{equation}
In the amplitude, the part $\frac{4\mu_0}{(M_{J/\psi}^2-t)(2\mu_0^2+M_{J/\psi}^2-t)}$ are from the propagators in the quark loop of the $\gamma\mathcal{P}J/\psi$ vertex, where the off-shell effect are included by a form factor $\mu_0^2/(\mu_0^2+p^2)$ with $\mu_0=1.1~{\rm GeV}$; $u(p_2)$ and $u(p_4)$ are the wave function of the proton in the initial and final states, respectively; $\beta_c$ is a parameter which represents the coupling strength between the Pomeron and the charm quark. The values of the parameters employed in the calculation are discussed in the Appendix. A more detailed introduction of the Pomeron exchange model for vector meson photoproductions can be found in Refs.~\cite{Pichowsky:1996jx, Pichowsky:1996tn, Laget:1994ba, Lee:2022ymp, Zhao:1999af, Donnachie:1988nj}.

\subsection{$s$-channel transitions}
As shown in Fig.~\ref{f1}(b, c), the $s-$channel scattering are expressed by the loop transitions with intermediate $\Lambda_c\bar{D}^{(*)}$ pairs and by the tree-level diagrams with the intermediate $P_c$ resonances. Although the $P_c$ poles are included directly as a tree-level contribution, one recognizes that the $P_c$ states can be dynamically generated by the open-charm $\Sigma_c\bar{D}^{(*)}$ scatterings in the hadronic molecule picture. Because of this, it can be understood that the production of the $P_c$ states in $\gamma p\to J/\psi p$ is a subleading process.

In Fig.~\ref{f1}(b, c) the vector meson dominance (VMD) model~\cite{Kroll:1967it,Bauer:1977iq,OConnell:1995nse} is adopted for describing the photon coupling to the charmed meson pair and to the $P_c\to \gamma p$. The Lagrangian for the $\gamma V$ coupling is
\begin{equation}\label{eq7}
\mathcal{L}_{VMD}=-\frac{eM_{J/\psi}^2}{f_{J/\psi}}V^\mu A_\mu,\\
\end{equation}
where the $f_{J/\psi}$=11.18 is the decay constant of $J/\psi$ determined in $J/\psi\to e^+e^-$ channel, and $V^\mu$ and $A_\mu$ are the field operators of $J/\psi$ and photon, respectively. The Lagrangians for the $P_cJ/\psi p$ and $\Lambda_c\bar{D}^{(*)}J/\psi p$ couplings are as follows:
\begin{equation}\label{eq8}
\begin{split}
\mathcal{L}_{P_c(1/2^-)J/\psi p}=&g_{P_c(1/2^-)}\bar{N}\gamma_5\gamma_\rho(-g^{\rho\mu}+\frac{p^\rho p^\mu}{m_{P_c}^2})P_c\psi_\mu,\\
\mathcal{L}_{P_c(3/2^-)J/\psi p}=&g_{P_c(3/2^-)}\bar{N}P_c^\mu\psi_\mu,\\
\mathcal{L}_{J/\psi p\Lambda_c \bar{D}^{(*)}}&=ig_x\psi^\nu \bar{N}\gamma_5\gamma_\mu\Lambda_c\bar{D}+g_{x^*}\bar{N}\Lambda_c\psi^\nu D_\mu^*.\\
\end{split}
\end{equation}
With the above Lagrangians, the amplitudes of Fig.~\ref{f1}(b, c) can be written as:
\begin{equation}\label{eq9}
\begin{split}
\mathcal{M}^{P_c(4312)}&=-\frac{eM_{J/\psi}^2}{f_{J/\psi}}\bar{u}_p(p_4,m_4)\gamma_5\tilde{\gamma}_\mu[(\slashed{p}_1+\slashed{p}_2)+m_{P_c(4312)}]\\
&\times\tilde{\gamma}_\nu\gamma_5u_p(p_2,m_2)\varepsilon_{J/\psi}^{*\mu}(-g_{\nu\alpha}+\frac{p_{1\nu}p_{1\alpha}}{m_{J/\psi}^2})\varepsilon_{\gamma\alpha}\\
&\times\frac{g_{P_c(4312)}^2}{((p_1+p_2)^2-m_{P_c(4312)}^2)(p_1^2-m_{J/\psi}^2)},\\
\mathcal{M}^{P_c(4440)}&=-\frac{eM_{J/\psi}^2}{f_{J/\psi}}\bar{u}_p(p_4,m_4)\gamma_5\tilde{\gamma}_\mu[(\slashed{p}_1+\slashed{p}_2)+m_{P_c(4440)}]\\
&\times\tilde{\gamma}_\nu\gamma_5u_p(p_2,m_2)\varepsilon_{J/\psi}^{*\mu}(-g_{\nu\alpha}+\frac{p_{1\nu}p_{1\alpha}}{m_{J/\psi}^2})\varepsilon_{\gamma\alpha}\\
&\times\frac{g_{P_c(4440)}^2}{((p_1+p_2)^2-m_{P_c(4440)}^2)(p_1^2-m_{J/\psi}^2)},\\
\mathcal{M}^{P_c(4380)}&=-\frac{eM_{J/\psi}^2}{f_{J/\psi}}\bar{u}_p(p_4,m_4)[(\slashed{p}_1+\slashed{p}_2)+m_{P_c(4380)}]\\
&\times[-g_{\mu\nu}+\frac{1}{3}\gamma_\mu\gamma_\nu+\frac{1}{3}\frac{\slashed{q}}{q^2}(\gamma_\mu q_\nu-\gamma_\nu q_\mu)+\frac{2}{3}\frac{q_\mu q_\nu}{q^2}]\\
&\times u_p(p_2,m_2)\varepsilon_{J/\psi}^{*\mu}(-g^{\nu\alpha}+\frac{p_1^nu p_1^\alpha}{m_1^2})\varepsilon_{\gamma\alpha}\\
&\times\frac{g_{P_c(4380)}^2}{((p_1+p_2)^2-m_{P_c(4380)}^2)(p_1^2-m_{J/\psi}^2)},\\
\mathcal{M}^{P_c(4457)}&=-\frac{eM_{J/\psi}^2}{f_{J/\psi}}\bar{u}_p(p_4,m_4)[(\slashed{p}_1+\slashed{p}_2)+m_{P_c(4457)}]\\
&\times[-g_{\mu\nu}+\frac{1}{3}\gamma_\mu\gamma_\nu+\frac{1}{3}\frac{\slashed{q}}{q^2}(\gamma_\mu q_\nu-\gamma_\nu q_\mu)+\frac{2}{3}\frac{q_\mu q_\nu}{q^2}]\\
&\times u_p(p_2,m_2)\varepsilon_{J/\psi}^{*\mu}(-g^{\nu\alpha}+\frac{p_1^nu p_1^\alpha}{m_1^2})\varepsilon_{\gamma\alpha}\\
&\times\frac{g_{P_c(4457)}^2}{((p_1+p_2)^2-m_{P_c(4457)}^2)(p_1^2-m_{J/\psi}^2)},\\
\mathcal{M}_{\Lambda_c\bar{D}}&=\int \frac{dq_1^4}{(2\pi)^4}ig_x^2\frac{eM_{J/\psi}^2}{f_{J/\psi}}\bar{u}_p(p_4,m_4)\gamma_5\gamma_\mu(\slashed{q}_1+m_1)\\
&\times\gamma_\nu\gamma_5u_p(p_2,m_2)\varepsilon_{J/\psi}^{*\mu}(p_3,m_3)\varepsilon_{\gamma\alpha}(p_1,m_1)\\
&\times\frac{g^{\nu\alpha}\mathcal{F}^2(q_1^2, \Lambda^2)}{(q_1^2-m_{q_1}^2)(q_2^2-m_{q_2}^2)(p_1^2-m_{J/\psi}^2)},\\
\mathcal{M}_{\Lambda_c\bar{D}^*}&=\int
\frac{dq_1^4}{(2\pi)^4}ig_x^2\frac{eM_{J/\psi}^2}{f_{J/\psi}}\bar{u}_p(p_4,m_4)(\slashed{q}_1+m_{q_1})u_p(p_2,m_2)\\
&\times(-g^{\mu\nu}+\frac{q_2^\mu q_2^\nu}{m_{D^*}^2})\varepsilon_{J/\psi\mu}^{*}(p_3,m_3)\varepsilon_\gamma^\alpha(p_1,m_1)\\
&\times\frac{g_{\nu\alpha}\mathcal{F}^2(q_1^2, \Lambda^2)}{(q_1^2-m_{q_1}^2)(q_2^2-m_{q_2}^2)(p_1^2-m_{J/\psi}^2)}.\\
\end{split}
\end{equation}
In the amplitudes, $\mathcal{F}^2(q_1^2, \Lambda^2)=e^{-2|\mathbf{q}_1|^2/\Lambda^2}$ is a form factor which takes into account the effects from the finite size of hadrons, and $\Lambda=0.5~{\rm GeV}$ is a typical value adopted. With the reduction of the tensor integral in the above amplitudes, the loop integrals can be represented by the Passarino-Veltman scalar integrals \cite{tHooft:1978jhc}. Finally, the analytic expression of the scalar integral with the form factor can be written as~\cite{Gong:2022hgd, Guo:2014iya, Cao:2017lui}:
\begin{equation}
\small
\begin{split}
&\int \frac{e^{-\frac{2|\mathbf{q}_1|^2}{\Lambda^2}}}{(q_1^2-m_{q_1}^2+i\varepsilon)(q_2^2-m_{q_2}^2+i\varepsilon)}dq_1^4\\
\approx&\frac{1}{4m_{q_1}m_{q_2}}\int dq_1^0d\mathbf{q}_1^3\frac{e^{-\frac{2|\mathbf{q}_1|^2}{\Lambda^2}}}{(q_1^0-m_{q_1}-\frac{|\mathbf{q}_1|^2}{2m_{q_1}}+i\varepsilon)(\sqrt{s}-q_1^0-m_{q_2}-\frac{|\mathbf{q}_2|^2}{2m_{q_2}}+i\varepsilon)}\\
=&\frac{2\pi i}{4m_{q_1}m_{q_2}}\int d\mathbf{q}_1^3\frac{e^{-\frac{2|\mathbf{q}_1|^2}{\Lambda^2}}}{\sqrt{s}-m_{q_1}-m_{q_2}-\frac{|\mathbf{q}_1|^2}{2m_{q_1}}-\frac{|\mathbf{q}_2|^2}{2m_{q_2}}}\\
=&\frac{i(2\pi)^3}{4m_{q_1}m_{q_2}}[\frac{\mu\Lambda}{\sqrt{2\pi}}+\mu ke^{-2k^2/\Lambda^2}(-erfi[\frac{\sqrt{2}k}{\Lambda}]+i)].\\
\end{split}
\end{equation}
In the above formula, $\mu=\frac{m_{q_1}m_{q_2}}{m_{q_1}+m_{q_2}}$ is the reduced mass. $k$ is determined as $\frac{k^2}{2\mu}=\sqrt{s}-m_{q_1}-m_{q_2}$. With the exponential form factor, the artificial pole involved in the denominator of the monopole form factor is avoid. As a result, the unphysical peak induced by the artificial pole will not appear in the cross section.

Now, with the amplitudes shown in the above subsections, the total amplitude can be expressed as
\begin{equation}\label{eq11}
\begin{split}
\mathcal{M_{\rm tot}}=&\mathcal{M^P}+\mathcal{M}^{P_c(4312)}+\mathcal{M}^{P_c(4380)}+\mathcal{M}^{P_c(4440)}\\
&+\mathcal{M}^{P_c(4457)}+\mathcal{M}_{\Lambda_c\bar{D}}+\mathcal{M}_{\Lambda_c\bar{D}^*}.\\
\end{split}
\end{equation}
With the VMD model for the photon and $J/\psi$ coupling, we note that there is no additional phase angle to be introduced into the above equation among amplitudes for different mechanisms.

\subsection{Polarized beam asymmetry}
Since the spin polarization observables can always provide additional dynamic information about the transition mechanisms~\cite{Conzett:1994rg, Fasano:1992es, Pichowsky:1994gh, Zhao:2005vh}, the access to the polarized beam asymmetry at JLab would be extremely valuable for a better understanding of the transition mechanism for $\gamma p\to J/\psi p$ near threshold. We focus on the linearly polarized photon beam which is defined as:
\begin{equation}\label{eqBA}
\begin{split}
\check{\Sigma}=\frac{d\sigma_x-d\sigma_y}{d\sigma_x+d\sigma_y},\\
\end{split}
\end{equation}
where $d\sigma$ denotes the differential cross section. With the initial photon momentum defined as the $z-$axis, in the overall center of mass (c.m.) frame the linearly polarized photon can be along the $x$-axis which is within the reaction plane, or along the $y$-axis which is perpendicular to the the reaction plane. The sum of $d\sigma_x$ and $d\sigma_y$ gives the unpolarized differential cross section of the process. In order to calculate the differential cross section with the polarized photon beam, we need to calculate the helicity amplitudes of the process firstly. The helicity amplitudes are defined as:
\begin{equation}
\begin{split}
\mathcal{M}^{\lambda_\gamma\lambda_i\lambda_f\lambda_V}&=\langle\varepsilon_{J/\psi}^{\lambda_V}(p_3, m_3)\bar{u}_p^{\lambda_f}(p_4, m_4)|\hat{T}|u_p^{\lambda_i}(p_2, m_2)\varepsilon_{\gamma}^{\lambda_\gamma}(p_1)\rangle,\\
\end{split}
\end{equation}
where $\lambda_\gamma$, $\lambda_i$, $\lambda_f$, and $\lambda_V$ represent the corresponding helicities of the  photon, initial proton, final proton, and $J/\psi$, respectively. The operator $\hat{T}$ contains the dynamics introduced by different mechanisms.

In the helicity amplitudes, the polarization vector of the photon is defined as:
\begin{equation}
\begin{split}
\varepsilon_{\gamma}^{+1}&=\frac{-1}{\sqrt{2}}(0, 1, i, 0),~\varepsilon_{\gamma}^{-1}=\frac{1}{\sqrt{2}}(0, 1, -i, 0),\\
\varepsilon_{\gamma}^x&=-\frac{1}{\sqrt{2}}(\varepsilon_{\gamma}^{+1}-\varepsilon_{\gamma}^{-1})=(0, 1, 0, 0),\\
\varepsilon_{\gamma}^y&=\frac{i}{\sqrt{2}}(\varepsilon_{\gamma}^{+1}+\varepsilon_{\gamma}^{-1})=(0, 0, 1, 0).\\
\end{split}
\end{equation}
With above relations, the amplitudes in $d\sigma_x$ and $d\sigma_y$ can be obtained from the helicity amplitude like:
\begin{equation}
\begin{split}
\mathcal{M}_x&=\sum_{\lambda_{i,f}=\pm\frac{1}{2}}\sum_{\lambda_V=\pm1,0}\frac{-1}{\sqrt{2}}(\mathcal{M}^{\lambda_i\lambda_f\lambda_V}_{\lambda_\gamma=+1}-\mathcal{M}^{\lambda_i\lambda_f\lambda_V}_{\lambda_\gamma=-1}),\\
\mathcal{M}_y&=\sum_{\lambda_{i,f}=\pm\frac{1}{2}}\sum_{\lambda_V=\pm1,0}\frac{i}{\sqrt{2}}(\mathcal{M}^{\lambda_i\lambda_f\lambda_V}_{\lambda_\gamma=+1}+\mathcal{M}^{\lambda_i\lambda_f\lambda_V}_{\lambda_\gamma=-1}).\\
\end{split}
\end{equation}
With the above amplitudes, the corresponding differential cross sections of $\gamma p\to J/\psi p$ with the polarized photon beam can be determined. The polarized beam asymmetry can finally be expressed as
\begin{equation}\small\label{eq16}
\begin{split}
\check{\Sigma}&=\frac{1}{2|\mathcal{M}_{\rm tot}|^2}(-\mathcal{M}^{1,-1}_r\mathcal{M}^{4,+1}_r-\mathcal{M}^{1,-1}_i\mathcal{M}^{4,+1}_i+\mathcal{M}^{1,0}_r\mathcal{M}^{4,0}_r+\mathcal{M}^{1,0}_i\mathcal{M}^{4,0}_i\\
&-\mathcal{M}^{1,+1}_r\mathcal{M}^{4,-1}_r-\mathcal{M}^{1,+1}_i\mathcal{M}^{4,-1}_i+\mathcal{M}^{2,-1}_r\mathcal{M}^{3,+1}_r+\mathcal{M}^{2,-1}_i\mathcal{M}^{3,+1}_i\\
&-\mathcal{M}^{2,0}_r\mathcal{M}^{3,0}_r-\mathcal{M}^{2,0}_i\mathcal{M}^{3,0}_i+\mathcal{M}^{2,1}_r\mathcal{M}^{3,-1}_r+\mathcal{M}^{2,1}_i\mathcal{M}^{3,-1}_i).\\
\end{split}
\end{equation}
Here, the subscript $r(i)$ means the real(imaginary) part of an amplitude. The abbreviations $\mathcal{M}^{1,\lambda_V}=\mathcal{M}^{1,-\frac{1}{2},+\frac{1}{2},\lambda_V}$, $\mathcal{M}^{2,\lambda_V}=\mathcal{M}^{1,+\frac{1}{2},+\frac{1}{2},\lambda_V}$,
$\mathcal{M}^{3,\lambda_V}=\mathcal{M}^{1,-\frac{1}{2},-\frac{1}{2},\lambda_V}$, and $\mathcal{M}^{4,\lambda_V}=\mathcal{M}^{1,+\frac{1}{2},-\frac{1}{2},\lambda_V}$ are adopted.

%*********************************************************************
\section{result and discussion}\label{sec3}
%*********************************************************************
In our model the undetermined parameters include the $\Lambda_c\bar{D}^{(*)}J/\psi p$ and $P_c J/\psi p$ couplings. The former one is determined in Ref.~\cite{Duan:2023dky, Khodjamirian:2011jp, Khodjamirian:2011sp, Lin:2019qiv, Lin:2017mtz, Du:2020bqj}. The latter one is adopted from the $P_c$ decays into $J/\psi p$ in the hadronic molecule picture.

\subsection{Total and differential cross sections}
\begin{center}
\begin{figure*}[ht]
\includegraphics[width=16.8cm,height=8cm]{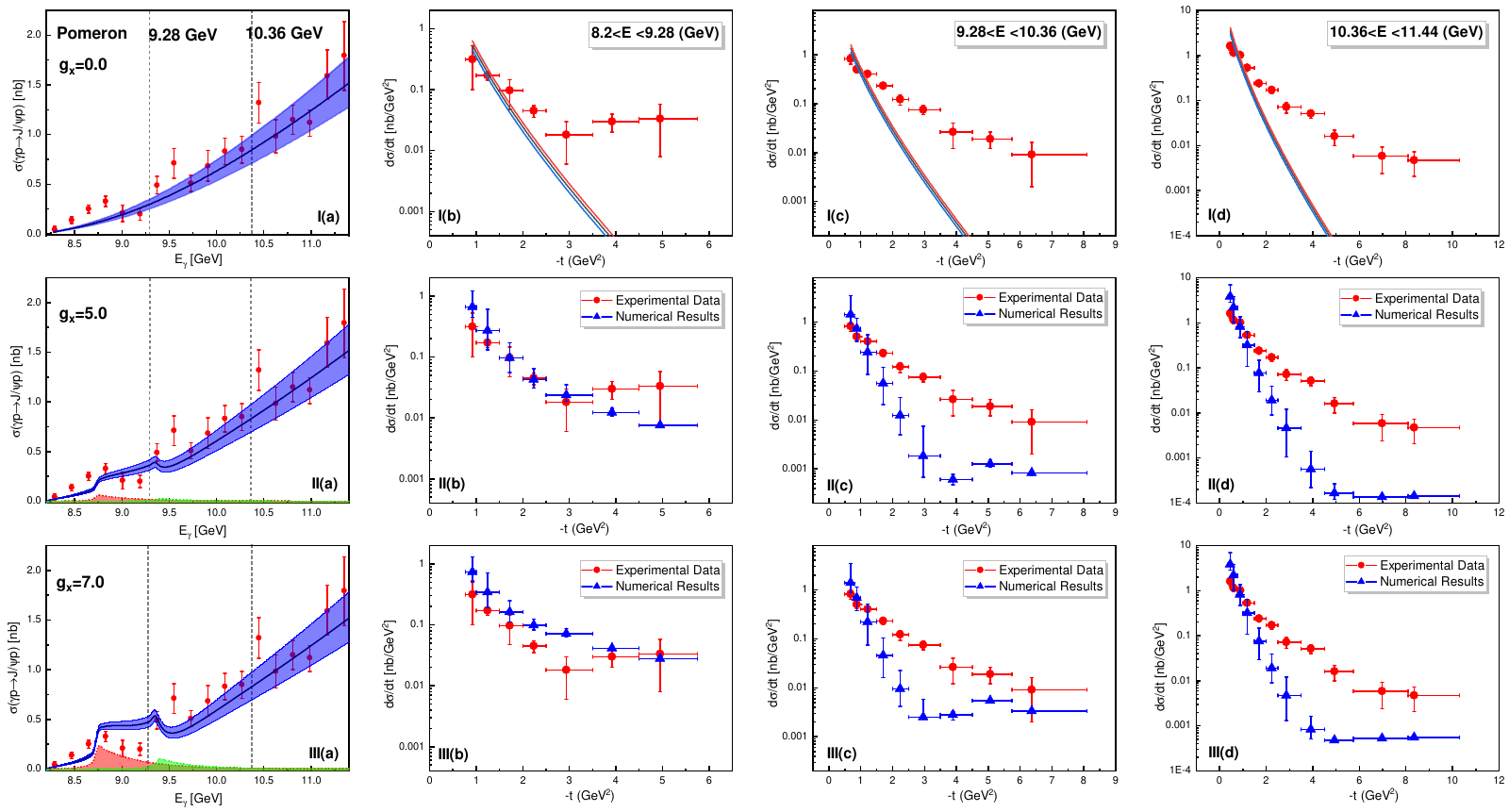}
\caption{The total and differential cross sections from the Pomeron exchange and $\Lambda_c\bar{D}^{(*)}$ loop transitions. The graphs with subscripts (a), (b), (c), and (d) represent the total cross section, the differential cross section in $8.2~{\rm GeV}<E_{\gamma}<9.28~{\rm GeV}$, $9.28~{\rm GeV}<E_{\gamma}<10.36~{\rm GeV}$, and $10.36~{\rm GeV}<E_{\gamma}<11.44~{\rm GeV}$, respectively. The graphs with subscripts I, II, and III represent the results from Pomeron exchange, the Pomeron exchange with additional $\Lambda_c\bar{D}$ loop in $g_x=5$ GeV$^{-1}$ and $g_x=7$ GeV$^{-1}$, respectively. The red and green peaks in the total cross section represent the individual contributions from the $\Lambda_c\bar{D}^{(*)}$ loops.}
\label{f3}
\end{figure*}
\end{center}

Proceeding to the numerical results, our calculations are compared with the GlueX data~\cite{GlueX:2023pev}. The total cross sections, differential cross sections collected at $8.2~{\rm GeV}<E_{\gamma}<9.28~{\rm GeV}$, $9.28~{\rm GeV}<E_{\gamma}<10.36~{\rm GeV}$, and $10.36~{\rm GeV}<E_{\gamma}<11.44~{\rm GeV}$ are shown in the four columns from left to right, respectively, in Fig.~\ref{f3}. The data are shown by the (red) solid dots with errors. Meanwhile, we present three sets of calculations to compare with the data in three rows. In the first row the exclusive contributions from the Pomeron exchange is illustrated by the shadowed bands, i.e. with $g_x=0$ and $0.22~{\rm GeV^{-1}}<\beta_c<0.26~{\rm GeV^{-1}}$. The range of $\beta_c$ is determined by the high-energy experiment as discussed in the Appendix. The solid line denotes the results with $\beta_c=0.24 \ {\rm GeV^{-1}}$. It shows the dominance of the Pomeron exchange contributions has almost saturated the total cross sections. However, the Pomeron exchange contributions only account for the diffractive transitions at small $|t|$, and it is clear that the differential cross sections at large $|t|$ (i.e. large scattering angles) have significant deficits.

In the second row the open-charm $\Lambda_c\bar{D}^{(*)}$ contributions are included with a strength of $g_x=5$ GeV$^{-1}$, and in the third row with $g_x=7$ GeV$^{-1}$. These are the values which can provide a reasonable contribution of the $\Lambda_c\bar{D}^{(*)}$ loop compared with the experimental data. Meanwhile, the value of $g_x=5$ GeV$^{-1}$ seems to be the limit that can be accommodated by the experimental data. It is slightly larger than that determined in our previous work~\cite{Duan:2023dky}, i.e. $|g_x|=3$ GeV$^{-1}$, but can still be regarded as consistent. For the total cross section the theoretical results are shown by the shadowed band which is determined by the range of $\beta_c$ in the Pomeron exchange. Since the Pomeron exchange plays a dominant role, its uncertainties have largely determined the behavior of the total cross sections. With different couplings for the open-charm channels, it shows that structures can arise from the $\Lambda_c\bar{D}^{(*)}$ thresholds. In contrast, the $s$-channel processes are generally small, but may produce measurable effects in the large scattering angle region. Note that the experimental data for the differential cross sections are collected at three energy ranges, i.e. $8.2~{\rm GeV}<E_{\gamma}<9.28~{\rm GeV}$, $9.28~{\rm GeV}<E_{\gamma}<10.36~{\rm GeV}$, and $10.36~{\rm GeV}<E_{\gamma}<11.44~{\rm GeV}$. It means that the differential cross sections are averaged within certain ranges of $-t$. To compare with the data, we divide the differential cross section into different $-t$ bins according to the experimental analysis for each energy range of $E_\gamma$. Within each $-t$ bin the differential cross section is calculated and averaged. The vertical errors denote the range of differential cross sections within the $-t$ bin and the central value is the averaged one. It should be noted that we have fixed $\beta_c=0.24$ GeV$^{-1}$ in the differential cross sections in the second and third rows.

In the second and third row of Fig.~\ref{f3} the calculated and averaged differential cross sections for each $-t$ bin and each energy region are presented by the solid triangular dots with errors. It shows that the differential cross sections in the energy region of  $8.2~{\rm GeV}<E_{\gamma}<9.28~{\rm GeV}$ can be well described by taking into account the $\Lambda_c\bar{D}^{(*)}$ open-charm contributions. The data show an apparent enhancement at the large scattering angles, which can be reasonably described by the $\Lambda_c\bar{D}^{(*)}$ open-charm contributions with $g_x=5$ or 7 GeV$^{-1}$. In contrast, the inclusion of the $\Lambda_c\bar{D}^{(*)}$ open-charm contributions seems not to be sufficient for describing the data for higher energy regions. It suggests that additional mechanisms should play a role, and it naturally calls for the inclusion of the $P_c$ states.

\begin{center}
\begin{figure*}[htbp]
\includegraphics[width=16.4cm,height=8cm]{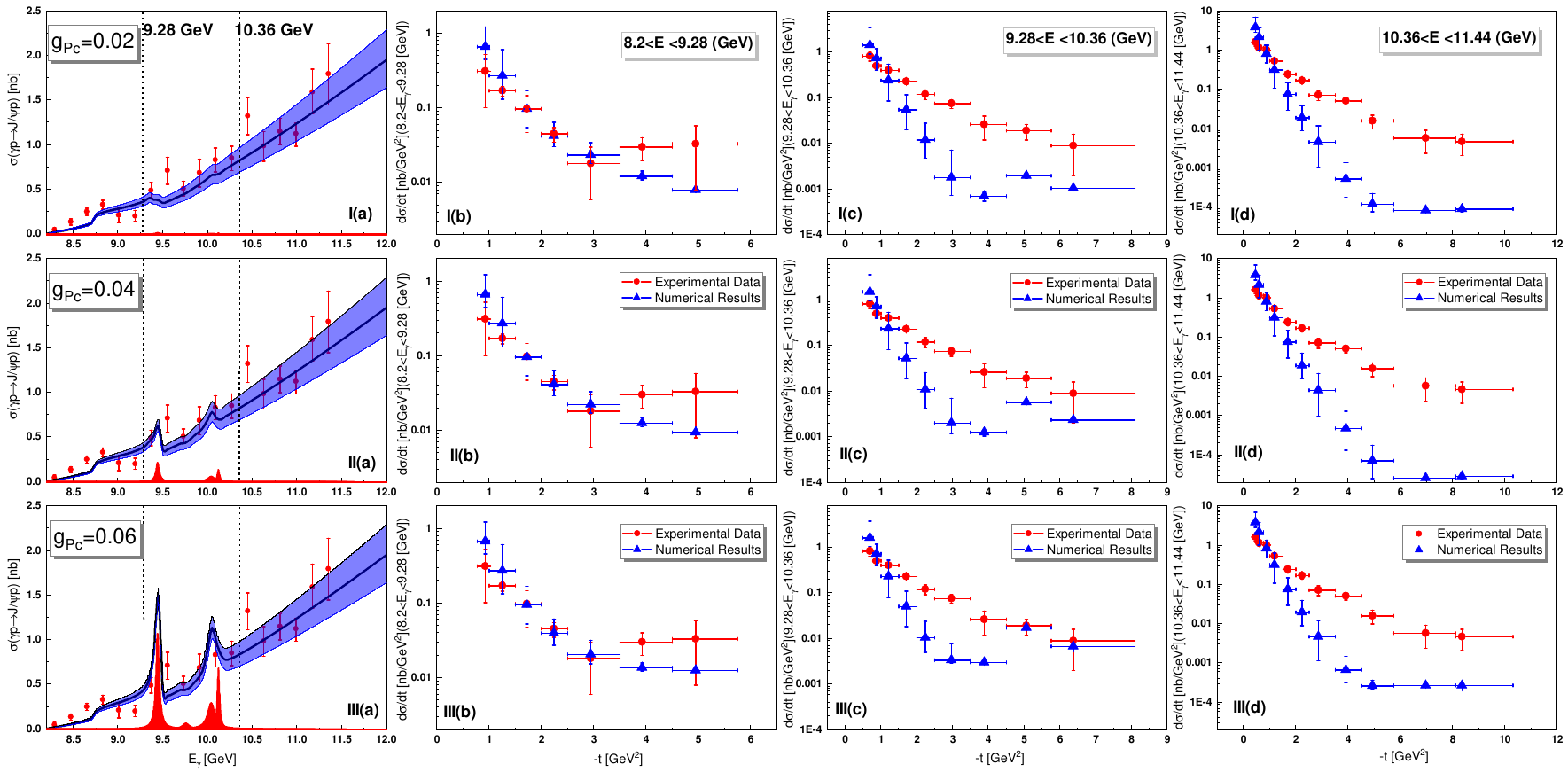}
\caption{The total and differential cross sections from the intermediate $P_c$ resonance and the additional Pomeron exchange and $\Lambda_c\bar{D}^{(*)}$ loops. The graphs with subscripts (a), (b), (c), and (d) represent the total cross section, the differential cross section in $8.2~{\rm GeV}<E_{\gamma}<9.28~{\rm GeV}$, $9.28~{\rm GeV}<E_{\gamma}<10.36~{\rm GeV}$, and $10.36~{\rm GeV}<E_{\gamma}<11.44~{\rm GeV}$, respectively. The graphs with subscripts I, II, and III represent the results with $g_{P_c}=0.02$, $g_{P_c}=0.04$, and $g_{P_c}=0.06$. The red peaks in the total cross section represent the individual contributions from different $P_c$ states. }
\label{f4}
\end{figure*}
\end{center}

Naturally, the $P_c$ states will contribute in the higher energy region. As the hadronic molecules formed by the $S$-wave $\Sigma_c^{(*)}\bar{D}^{(*)}$ interactions, their couplings to the $J/\psi p$ channel can be constructed by the $\Sigma_c^{(*)}\bar{D}^{(*)}$ rescatterings into $J/\psi p$~\cite{Lin:2017mtz, Lin:2019qiv}. Based on the heavy quark symmetry and heavy quark spin symmetry, these couplings are connected by an overall coupling $g_{P_c}$, which in principle can be constrained by experimental data. However, since the present experimental data still cannot provide a strong constraint on the pole positions, the determined coupling $g_{P_c}$ also has a range of uncertainties. In Table~\ref{Tab1} we list different values of $g_{P_c}$ with which the $P_c$ couplings to $J/\psi p$ can be determined. In Fig.~\ref{f4} with the Pomeron exchange ($0.22~{\rm GeV^{-1}}<\beta_c<0.26~{\rm GeV^{-1}}$) and  $\Lambda_c\bar{D}^{(*)}$ open-charm channel ($g_x=5$ GeV$^{-1}$) contributions, the total and differential cross sections are presented for different $g_{P_c}$ values, namely, $g_{P_c}=0.2, \ 0.4, \ 0.6$, in three rows, respectively. The varying range of $g_{P_c}$ reflects the uncertainties arising from the $P_c$ productions, while the range of $\beta_c$ reflects the uncertainties from the Pomeron exchange contributions.

As shown by Fig.~\ref{f4}, the differential cross section data at $8.2~{\rm GeV}<E_{\gamma}<9.28~{\rm GeV}$ can still be well described, while the data at $9.28~{\rm GeV}<E_{\gamma}<10.36~{\rm GeV}$ get significantly improved. In contrast, the calculated results at $10.36~{\rm GeV}<E_{\gamma}<11.44~{\rm GeV}$ turn out to be still insufficient to account for the data at the large scattering angles. Meanwhile, the total cross sections can accommodate the $P_c$ contributions with a reasonable range of the coupling $g_{P_c}$. As shown in Table~\ref{Tab1}, the partial widths for the decay of these $P_c$ states into $J/\psi p$ are consistent with the experimental analysis. It means that the production of $P_c$ states in the $J/\psi$ photoproduction indeed experiences a double suppression at the production and decay vertices. This can be regarded as a consequence of the hadronic molecule picture for these $P_c$ states. It should also be addressed that the deficit in the differential cross sections at higher energies is an indication for further $s$-channel processes to contribute. For instance, open-charm thresholds such as $\Lambda_c^*\bar{D}^*$ and $\Sigma_c^*\bar{D}^*$ will be present and play a role. Although the total cross section is not expected to change a lot in the magnitude, they will produce non-negligible effects in the differential cross section at the large scattering angles.

\begin{table}[htbp]
\caption{The partial decay width in the $J/\psi p$ channel of the $P_c$ states. }
\label{Tab1}
\renewcommand\arraystretch{1.20}
\begin{tabular*}{86mm}{@{\extracolsep{\fill}}ccccc}
\toprule[1.0pt]
\toprule[1.0pt]
$g_{P_c}$   &0.01   &0.02   &0.04   &0.06   \\
\toprule[0.8pt]
$\Gamma[P_c(4312)\to J/\psi p]$ (keV)&7.71&30.8&123&277\\
$\Gamma[P_c(4380)\to J/\psi p]$ (keV)&2.93&11.7&46.9&105\\
$\Gamma[P_c(4440)\to J/\psi p]$ (keV)&9.69&38.8&155&349\\
$\Gamma[P_c(4457)\to J/\psi p]$ (keV)&3.31&13.3&53.0&119\\
\bottomrule[1.0pt]
\bottomrule[1.0pt]
\end{tabular*}
\end{table}

\subsection{Polarized beam asymmetry}

The role played by the $s$-channel processes makes the measurement of the spin polarization observables extremely valuable. In this subsection we focus on the polarized beam asymmetry and demonstrate its sensitivity to the $s$-channel processes at the large scattering angles.

\begin{center}
\begin{figure}[htbp]
\includegraphics[width=8.8cm,height=4.5cm]{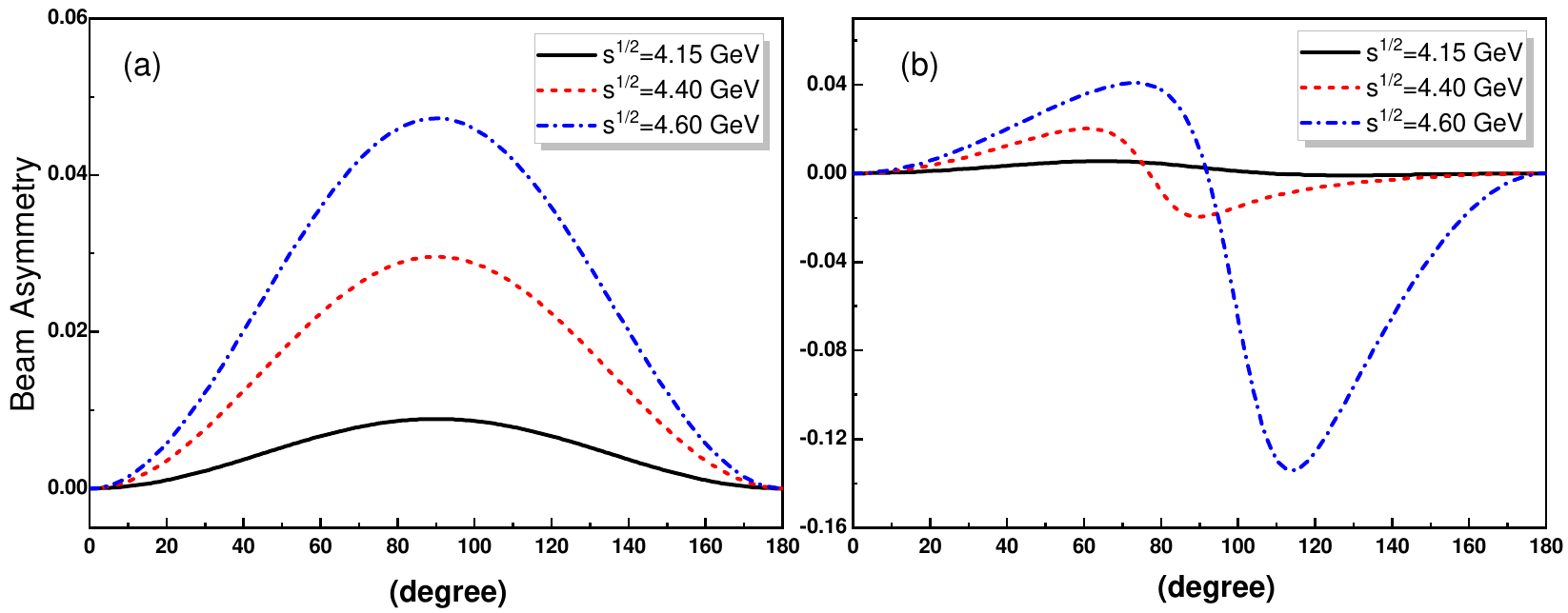}
\caption{The beam asymmetry predicted from (a) the Pomeron exchange, (b) the Pomeron exchange with $s-$channel contribution including the $\Lambda_c\bar{D}^{(*)}$ loops and $P_c$ intermediate resonances.}
\label{f5}
\end{figure}
\end{center}

In Fig.~\ref{f5}, the polarized beam asymmetry is plotted in terms of the scattering angle $\theta$ at three energies, i.e. $\sqrt{s}=4.15~{\rm GeV}$ (black solid line), $4.40~{\rm GeV}$ (red dotted line), and $4.60~{\rm GeV}$ (blue dot-dashed line). In Fig.~\ref{f5} (a) only the dominant diffractive contributions from the Pomeron exchange are calculated. It shows that non-trival structures arise at intermediate and large scattering angles. Recognizing that the Pomeron exchange drops quickly in terms of the c.m. energy, the large-angle asymmetries would not be reliable with the exclusive Pomeron exchange contribution. With the inclusion of the $s$-channel contributions, i.e. the open-charm $\Lambda_c \bar{D}^{(*)}$ and the $P_c$ states, the polarized beam asymmetries for these three energies are plotted in In Fig.~\ref{f5} (b) in terms of the scattering angle $\theta$. It turns out that significant interfering effects among these amplitudes can occur in the polarized beam asymmetry. In particular, the large-angle asymmetries are sensitive to the presence of the $s$-channel contributions. One sees that at the energies of $\sqrt{s}=4.40~{\rm GeV}$ (red dotted line), and $4.60~{\rm GeV}$ (blue dot-dashed line) there appears a node structure in the beam asymmetry.  This indicates the importance of this observable in disentangling the underlying dynamics, and experimental measurement of this observable will be valuable for further understanding of the reaction mechanism.

%**********************************************
\section{Summary}\label{sec4}
%**********************************************
The photoproduction process $\gamma p\to J/\psi p$ can provide a good opportunity for studying the open-charm channels and the nature of the hidden-charm pentaquark $P_c$ states. Although the recent experimental data from the GlueX Collaboration do not show clear signals for the $P_c$ production in the $s$ channel, we find that the evidence for the $\Lambda_c\bar{D}^{(*)}$ open-charm effects can provide a reasonable constraint on the significant of these $P_c$ signals. Namely, in $\gamma p\to J/\psi p$ the production of the hidden-charm $P_c$ states is doubly-suppressed by their couplings to the $\gamma p$ and $J/\psi p$ channels which are through the intermediate $\Sigma_c\bar{D}^{(*)}$ rescatterings. Our calculation shows that the hadronic molecule picture for the hidden-charm pentaquark $P_c$ states is not in conflict with the GlueX observation.

In order to better understand the property of these $P_c$ states, we propose experimental measurement of the spin polarization observables. In particular, we show that the polarized beam asymmetry is sensitive to the presence of the $s$-channel processes, and future experimental measurement of this observable at JLab will provide valuable information about the $s$-channel dynamics.

%******************************************************************
\section*{Acknowledgements}
%******************************************************************

This work is supported, in part, by the National Natural Science Foundation of China (Grant No. 12347135 and 12235018), China Postdoctoral Science Foundation No. 2023M733502, DFG and NSFC funds to the Sino-German CRC 110 ``Symmetries and the Emergence of Structure in QCD" (NSFC Grant No. 12070131001, DFG Project-ID 196253076), National Key Basic Research Program of China under Contract No. 2020YFA0406300, and Strategic Priority Research Program of Chinese Academy of Sciences (Grant No. XDB34030302).

%******************************************************************
\section*{Appendix}
%******************************************************************

\begin{center}
\begin{figure}[htbp]
\includegraphics[width=8.2cm,height=7cm]{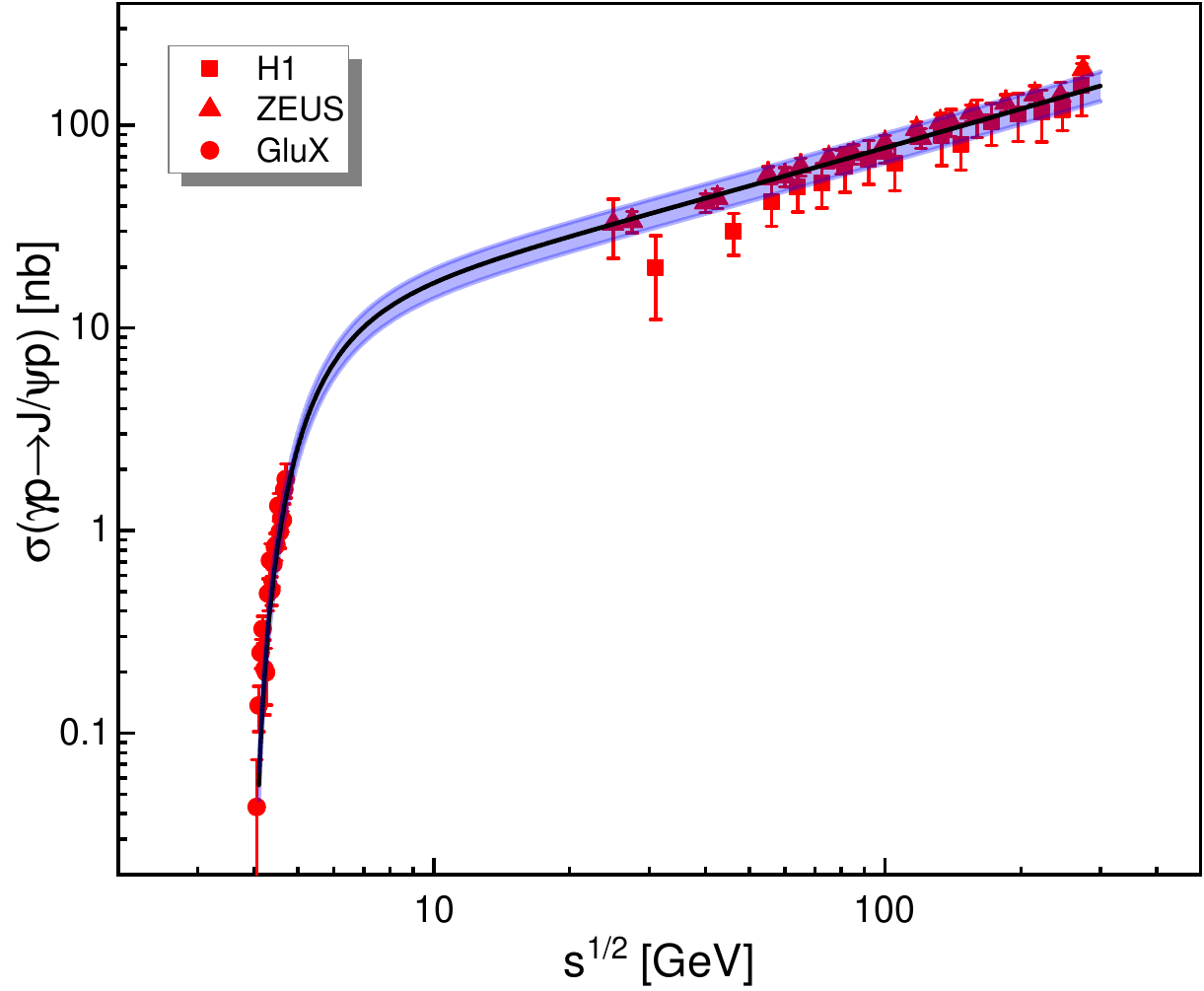}
\caption{The total cross sections for $\gamma p\to J/\psi p$ calculated by the Pomeron exchange model (shadowed band) to compared with the experimental data from the H1 Collaboration, ZEUS Collaboration, and the GlueX Collaboration~\cite{H1:2000kis, ZEUS:2002wfj, GlueX:2023pev}.
%The data from ZEUS include the $J/\psi$ reconstructed in $e^+e^-$ and $\mu^+\mu^-$ channels.
}
\label{fappendix}
\end{figure}
\end{center}

The parameters employed in the study of the Pomeron exchange are extracted from the photoproduction data, which should be explained here. In Fig.~\ref{fappendix}, we show the experimental data covering the energy range $4~{\rm GeV}<\sqrt{s}<300~{\rm GeV}$ from the H1 Collaboration, ZEUS Collaboration, and the GlueX Collaboration~\cite{H1:2000kis, ZEUS:2002wfj, GlueX:2023pev}. The parameters of the Pomeron exchange model are fitted by these data which means that the Pomeron exchange mechanism is extrapolated to the threshold region with a reasonable strength. We find that $\epsilon$ can be determined as $\epsilon=0.19$, which is slightly different from the value given in the Ref.~\cite{Lee:2022ymp}. To indicate the possible uncertainties within the Pomeron model, we also show the numerical results for a range of the Pomeron coupling to the constituent quark, i.e. $0.22~{\rm GeV^{-1}}<\beta_c<0.26~{\rm GeV^{-1}}$, which is illustrated by the shadowed band in Fig.~\ref{fappendix}. The solid line on top of the band  corresponds to the result with the central value $\beta_c=0.24~{\rm GeV^{-1}}$.

In Table~\ref{TabAppx} we calculate $\chi^2/ndf$ to obtain a rough estimate of the fitting quality with different parameters. In the two left columns $\chi^2/ndf$ is calculated with parameter $\epsilon$ fixed at 0.19 while $\beta_c$ varies with three different values. In the two right columns parameter $\beta_c$ is fixed at 0.24 ${\rm GeV^{-1}}$ while different values for $\epsilon$ are adopted.

\begin{table}[htbp]
\caption{The $\chi^2/ndf$ determined from the different parameters. $ndf$ is the numbers of the degree of the freedom.}
\label{TabAppx}
\renewcommand\arraystretch{1.20}
\begin{tabular*}{75mm}{@{\extracolsep{\fill}}cc|cc}
\toprule[1.0pt]
\toprule[1.0pt]
$\beta_c~[\rm GeV^{-1}]$&$\chi^2/ndf$&$\epsilon$&$\chi^2/ndf$  \\
\toprule[0.8pt]
0.22&$0.74\pm0.12$&0.17&$1.08\pm0.14$\\
0.24&$0.37\pm0.08$&0.19&$0.37\pm0.08$\\
0.26&$0.59^{+0.10}_{-0.11}$&0.21&$1.43\pm0.16$\\
\bottomrule[1.0pt]
\bottomrule[1.0pt]
\end{tabular*}
\end{table}

%******************************************************************

%******************************************************************
\end{document}